# TALENT MANAGEMENT – AN ETYMOLOGICAL STUDY

Author: Kiril Dimitrov

*Abstract:* The current article unveils and analyzes important shades of meaning for the widely discussed term ‚talent management'. It not only grounds the outlined perspectives in incremental formulation and elaboration of this construct, but also is oriented to exploring the underlying reasons for the social actors, proposing new nuances. Thus, a mind map and a fish-bone diagram are constructed to depict effectively and efficiently the current state of development for talent management and make easier the realizations of future research endeavours in this field.

**INTRODUCTION**

Discussions around the operationalization of the term „human resource management" (HRM) have a long history, revealing a number of influencing factors as: changing business environment conditions (globalization, diversity, complexity, ambiguity, profitability through growth, technology, intellectual capital, continuous change, unpredictability, misbalances, instability, cultural traversion across national borders, etc.); assumed reactions by company management teams to the changes, occurring on labor markets (e.g. proactive versus reactive managerial behavior, measurement of the effects of implemented human resource strategies upon the overall performance of a target unit or a whole company, emergence of the international HRM); the advances of general management theory, sociology, psychology and other sciences, representing sources of numerous ideas and techniques in the field of HRM (for example contingency approach, organisation development postulates, behavioural theory, cultural studies, etc.); the energy resource crisis from 1970s that brought to intensification of competition among companies; the strive for achieving a sustainable competitive advantage by management teams, building organizations to last; the competition between the "old world" (UK) and "the new one" (USA); strong desire of increasing the prestige for the professions in the HRM sphere; formulation of a convenient abbreviation for everyday use; emphasizing definite activities in personnel management; the refusal of HRM units to perform certain traditional people management activities; adopting HRM as a bundle of professional practices; understanding it as a means of human manipulation; the evolution of the roles, assigned to HR departments in the organizations; the development of laws in the sphere of labour relations; the importance and frequency of emerging issues and challenges in the HRM sphere; linguistic specificity; perceiving HRM as a hologram, etc. (Dimitrov, 2009; Nakata, 2009; Ulrich, 1998).

Differentiated and/ or frequently joint influence of all these factors creates diverse essence, elements, processes and practices for HRM in organizations and sometimes imposes deeper shifts in the professed paradigms, i.e. the emergence of a new term for denoting the implementations of contemporary people management practices and techniques in the organizations – *talent management*. Its adoption may be at least twice associated with recommending or assigning of new roles to human resource managers, specialists and functions for their units by opinion leaders in the field in order to secure at least the survival of the companies or in the best case their undergoing through seemless changes of functioning business model(s) in order to conquer and sustain a leading market position, as follows:

- *Around the deviding line of the last two centuries* the roles may be characterized by (Lipiec, 2001; Ulrich, 1998): (a) partnering with senior and line management in strategy execution; (b) serving as an administrative expert; (c) servicing simultaneously employees, employers and other outside clients; (d) becoming a continuous transformation agent; (e) orientation to long-term activities. The end of the Cold war and increasing globalization may be defined as key marker events to the formation of this bundle from HR roles.

- *Around the deviding line of the first two decades of the 21st century* the roles may be described by (Ulrich, Younger, Brockbank, Ulrich, 2012; Harvey, 2013): (a) transposing external business trends and stakeholder expectations into internal HR practices and actions; (b) earning personal credibility and taking an active position on business performance; (c) becoming an effective and efficient manager of revolving tensions between talent and teamwork, individual ability and organization capability, personal competence and organizational culture; (d) innovating and integrating separate HRM events into cohesive solutions in order to achieve sustainable results; (e) adding new attributes to an HR as a change agent, i.e. connecting the past to the future, and anticipating and managing individual, initiative, and institutional change; (f) flawlessly processing administrative work by means of advanced technology while generating information for more strategic work; (g) focusing on acquiring, developing, aligning and assessing people whose capabilities, skills and knowledge is becoming increasingly scarce. The World financial and economic crisis and intensified co-mingling, hybridizing, morphing, and clashing of cultures may be defined as a key marker event to the formation of newer and/ or richer bundle of HR roles.

The aim of this article is to take a snap-shot at the current bundle of attributes, constituting the contemporary meaning of the term „talent mnagement" by identifying and critically analyzing stable and emerging nuances, perspectives, relations with close terms, criteria, processes, strategic moves, etc., applied by different social actors to satisfy their insatiable necessity of creation and maintenance of competitive advantage through their human resources. This aim has to be achieved by means of deliberate directing researcher's attention to: (a) widely shared and accepted (older) information, included in (text)books, composed by prominent figures in the field of management, and (b) comparatively new information that is available in scientific articles from electronic databases[1], representing or containing as a section a respective etymological study for the term „talent management".

## 1. The dawn of talent management

Two key perspectives of analysis mark the evolutionary process of initial formation for the term „talent management". By their impact on the proceeding of this phenomenon these may be classified as: (a) a direct one, introducing the respective new terms and practices (a military-oriented perspective), and (b) an indirect one, characterized by incremental changes in dominating management reflections on the role of human resources in the building and lasting of business organizations (a cultural perspective).

*The military-oriented perspective* seems to constitute the core nuance in the meaning of talent management, a term that initially was professed in public through deliverables of McKinsey Consulting Company by means of a descriptive expression - "the war for talent". In academic and professional literature sources certain ambiguity exists in relation with the original source and year of publication where the term was coined for the first time, but this hesitation does not affect the applied expression (see table 1).

Table 1. The identified bundle of initial publications for talent management

| Source | Year of publication | Proclaimer and/or applied expression |
|---|---|---|
| Chambers, E. G., Foulon, M., Handfield-Jones, H., Hankin, S. M., Michaels III, E. G., The war for talent, 1998 Number 3, The McKinsey Quarterly: The Online Journal of McKinsey & Co., available at: http://www.mckinseyquarterly.com/article_print.aspx?L2=18&L3=31&ar=305 29/, accessed on: 28.07.2015). | 1998 | McKinsey company, "the war for talent" |
| Harvey, P., 15 years post McKinsey's 'war for talent' – have HR won a battle but lost the war?, a Kallidus company presentation, 26th March 2013, available at: www.google.com, accessed on: 19.08.2015. | 1997 | McKinsey company, "the war for talent", but an exact source is not mentioned |
| Axelrod, B., Handfield-Jones, H. and Michaels, E. (2002) "A new game plan for C players", Harvard Business Review, January, 81-88. and Michaels, E., Handfield-Jones, H. and Axelrod, B. (2001) The War for Talent, Boston, Harvard Business School Press. in Collings, D.G. and Mellahi, K. (2009) "Strategic Talent Management: A review and research agenda", Human Resource Management Review, 19: 4, 304–313, available at: : http://search.ebscohost.com, accessed on: 28.07.2015. | 1997 | A group of McKinsey consultants, "the war for talent" |

---

[1] http://search.ebscohost.com; www.emeraldinsight.com; www.sciencedirect.com, etc.;

Table 1. The identified of bundle initial publications for talent management (cont'd)

| Source | Year of publication | Proclaimer and/or applied expression |
|---|---|---|
| Michaels, E., Handfield-Jones, H., Axelrod, B., The War for Talent, Harvard Business Press, 2001, ISBN 9781578514595, mentioned in Iles, P., Chuai, X. and Preece, D. (2010) 'Talent Management and HRM in Multinational companies in Beijing: Definitions, differences and drivers', Journal of World Business, 45 (2), pp.179-189, available at: http://tees.openrepository.com/tees/handle/10149/95254, accessed on: 05.08.2015. | 2001 | "the war for talent" |
| Michaels, E., Handfield-Jones, H. and Beth, A. (2001). The War for Talent, McKinsey & Company, Inc.in Iles, P., Preece, D. and Chuai, X. (2010) 'Talent management as a management fashion in HRD: towards a research agenda', Human Resource Development International, 13(2), pp.125-145, available at: http://tees.openrepository.com/tees/handle/10149/107373, accessed on:02.08.2015. | 1997 is mentioned inside | "the war for talent" |
| Michaels, E., Handfield-Jones, H. and Beth, A. (2001, p. Xii, The War for Talent. Boston, MA: Harvard Business School Press. in Hatum (2010), A., Next generation talent management. Talent Management to Survive Turmoil, 2010, PALGRAVE MACMILLAN, p. 10) define talent as „the sum of a person's abilities – his or her intrinsic gifts, skills, knowledge, experience, intelligence, judgment, attitude, character, and drive. It also includes his or her ability to learn and grow." | 2001 | „talent" |

The phase of a wider dissemination of „the war for talent" concept among researchers and practicioners may be associated with a key marker event (2001) – the initial public expression of a strive for elucidating the contradictory nature of the „war for talent" concept that is reported a bit later and brillinatly revealed by Armstrong (2012) who compares the opposing points of view within this sub-sphere[2]. The supporting position is incarnated in the company imperatives, expressed by Michaels, Handfield-Jones and Axelrod (2001) who advise the companies to pursue them in order to win the 'war for managerial talent'. The last citation shows a narrower orientation of talent management to a certain group of personnel – the decision-makers in the business organization. The opposite position, that is formulated by Pfeffer (2001), belittles the advantages for the companies from adopting such an aggressive approach by describing its devastating effects on organization performance due to increasing demotivation, higher turnover, frequent demonstrations of arrogant behaviors, weakened capabilities of listening and learning among the employees, predominant reliance on employee promotions from outside, and languished interests in design and implementation of new and better management practices and executing changes in a target company's culture in order to (re-)(ab-)solve business-related problems (see table 2).

---

[2] Earlier versions of Armstrong's book were not found in the university library and the available scientific electronic databases, so the time lag between the real moment of first-time perceiving the contradictory nature of the „war for talent" concept and the moment of the formation of shared meaning and public reporting of its contents remained unclear for the author of this deliverable, although logically it is asumed that such a time lag exists.

Table 2. Diverse attitudes to the utility of „war for talent" concept for the business organizations

| The pillars, proposed by Michaels, Handfield-Jones and Axelrod (2001) as defenders of the concept | The doubts, expressed by Pfeffer (2001) as a criticiser of this concept |
|---|---|
| 1. It brings about deliberate creation of a winning employee value proposition that is considered as the company's driver for unique talent attraction. | 1. It contributes to posing an invariable emphasis on individual performance thereby damaging team work. |
| 2. It generates the management ability of moving beyond recruiting hype in order to build a long-term recruiting strategy. | 2. It generates a tendency to glorify the talents of those outside the company and downplay the skills and abilities of insiders. |
| 3. It implies the use of job experience and mentoring in order to cultivate the potential in managers. | 3. It provides diverse treatment to different groups of employees, classified by demonstrated performance levels, i.e. those labelled as less able become less able to a great (some) extent because they are asked to do less and given fewer resources and training. |
| 4. It requires strenghening of the talent pool by investing in A players (top performers), developing B players (employees with potential) and acting decisively on C players (poor peformers). | 4. It stimulates the mangers to de-emphasize the fixing of the systemic, cultural and business issues that are invariably more important for enhancing performance. |
| 5. It calls for development of a pervasive mindset, i.e. a deep conviction shared by leaders throughout the company that competitive advantage comes from deploying better talent at all levels. | 5. It brings to the development of an elitist, arrogant attitude among dcision-makers (recollect the Enron case). |
| Sources: Pfeffer (2001); Michaels, Handfield-Jones, Axelrod (2001). | |

The last known phase in the elaboration of the „war for talent" concept may be marked by the emergence of two new discernible streams, related with waging the intense battles within the continuous war for talent (1998 – up to the moment) by the involved constituences (company's managers, working people and others). *The first one* is based on formulating and paying a balanced attention to a bundle of important talent related issues as their attraction, development, retention, discharge and turnover. Furthermore, a change in the dominating career self-management paradigm is observed among working people who heavily rely on inter-company mobility in spite of sustainable management retention efforts. For example Somaya and Williamson (2011, p.75) consider that „perhaps it is time to declare that the war for talent is over . . . talent has won!". The challenge for the contemporary succeeding companies seems to be the utilization of potential benefits, provided by departing and former employees that may be realized by maintenance of appropriate relationships in order to achieve greater client access, further human capital access and higher generation of goodwill (Somaya, Williamson, 2011).

*The second one* is summarized by Paula Harvey (2013) who discusses the advantages and shortcommings for the strategic option of „ceasing the fire" in the field of talent war, basing her analysis on the results of several surveys, conducted by leading global institutions to support her opinion that talent management still representes one of the greatest risks for the majority of the operating companies in the United Kingdom Great

Britain and North Ireland and worldwide, characterized by appropriate talent shortages in spite of higher unemployment rates, shattered beliefs in achieved effectiveness in the implemented talent management processes, obvious necessities of undertaking potential changes in pursued talent management strategies, inevitable clarification (re-formulation) of HR's role in the process, urgent need of taking great pains in related spheres as improving leadership development and strategic workforce planning in the business organizations and the confronted difficulties in coping with constant attacks on the groups of potential senior executives for the companies by certain constituencies. Finally, the consultant proposes rearmament with new „weapons to win the war" that may be outlined in four perspectives, as follows:

- *The cultural perspective.* It is oriented to undertaking an obligatory change in the dominating HR mantra that only some of the working people in the organization represent its biggest asset and deserve fighting for their presence in and engagement with it.

- *Acting in accordance with the principles of the new „employer - employee" contract*, characterized by a greater bargaining power for the talented individual. It means that: (a) companies need people, (b) talented people constitute the completive advantage, (c) better talent makes a huge difference, (d) talented people and jobs are scarce, (e) people are mobile and their commitment is short term, (f) talent demands much more than a competitive remuneration offer.

- *Following the practical approach of giving recepies for developing a talent mind-set by the company management*, outlined by at least several potential strategic moves for the companies in the sphere as: developing a winning employee proposition, re-building of the implemented recruitment strategy, adopting development activities across all the functions of the organization with an accent on human resources whose performance and potential should be differentiated, and supporting the succeeding ones.

- *Reliance on the problem-solving approach in management* for making continuous invention of new reasons, justifying the desire of talents to choose to contribute to a target company and stay with it at least for a certain time period.

*There exists a second, often neglected, perspective*, contributing to the emergence of talent management theory at a later stage that was realized at (sub-)unconscious level. That is why it may be labeled as ‚a cultural perspective'. It is subjected to the assumption that the formation of certain interests and attitudes by some social actors (researchers, consultants, managers) to explain the connection between human resource development and organizational effectiveness (see Schein, 1977) gradually brought to the surface important and creative people management thinking and practices, almost 20 years later on officially sent and combined in the web of the new term of ‚talent management'.

## 2.  Relatively stable nuances in the meaning of "talent management"

An initial impression on talent management may be created only through identification of a clear-cut definition of the term. *For this purpose Armstrong's opinion*

*(2012) is preferred in this article, bearing in mind that the experienced researcher has demonstrated persistency in his HRM-related endeavours to develop and upgrade 12 consecutive issues of a textbook in this field.* According to him talent management represents „the process of ensuring that the organization has the talented people it needs to attain its business goals. It involves the strategic management of the flow of talent through an organization by creating and maintaining a talent pipeline" (see p.256).

It would be a satisfactory solution for the curious reader, if the scientist does not unravel the lack of consensus among the social actors in relation with the applied approaches in the sub-sphere. The identified interval of these approaches has the lowest limit, denoted by management succession planning and/or management development activities and the highest limit that incarnates „a more comprehensive and integrated bundle", adding a sound reliance on growth from within, considering it an important element of the pursued business strategy, precise and clear determinations and realizations of timely updates for needed competencies and qualities that talented incumbents should possess, deliberate maintenance of well-defined career paths, paying heavy attention to coaching and mentoring interactions in the company and making no concessions to high performance requirements.

Another reason for the occurrence of potential misunderstandings in relation with talent management essence is due to the observed significant gap between what directions of interests dominate in theory and practice, concerning HRM (Pfeffer, Sutton, 1999). Furthermore, the dominating beliefs and assumptions of business environment impacts on company performance, and the efficiency and effectiveness of preferred management approaches, tools and techniques in order to mitigate the short-term effects and long-term consequences, transform threats into opportunities, conquer and maintain a sustainable competitive advantage, have evolved for the last 25 years. That is why it sounds at least plausible that traditional approaches to HRM have served the (international) companies at a satisfactory level during the previous century. But the call for talent management may be justified by the new characterisitcs of the business environment nowadays that imply the use of new and innovative approaches in the development and deployment of human resources (Caligiuri, 2006; Lengnick-Hall, Andrade, 2008).

*The reliance on these reasons serves as a logic ground for Armstrong's (2012; 2011) reviewing of different facets in talent management* (see table 3).

Table 3. The facets of talent management, outlined by Michael Armstrong

| **Specific facet** | **Source** |
|---|---|
| An effective means of doing strategic investments in perosnnel members who are considered as assets and a source of competitive advantage within the perspective of human capital management (HCM). | Armstrong (2012), p. 72 |
| An important activity in people (employee) resourcing process. | Armstrong (2012), p. 201; Boxall and Purcell (2010), p. 29 in Armstrong (2011), p. 6 |
| Classifying talent management elements to transformational (identification and development of people with talent) and transactional (recruitment, administration of learning and development events) ones. | Armstrong (2012), p. 201 |
| Observed collision of its meaning with HRM or human resource development. The common ground requires „the right people in the right job at the right time and managing the supply and development of people throughout the organization". | Iles, Preece, Chuai (2010), p.127 in Armstrong (2012), p. 256 |
| It adds to Human resource development „a selective focus on a small 'talented' section of the workforce (a 'talent pool')". | Iles, Preece, Chuai (2010), p.127 in Armstrong (2012), p. 256 |
| Putting an emphasis on succession and human resource planning in the company which requires the adherence to important talent management activities as: (a) performing of organizationally-focused competence development and (b) specific construction, management and upgrade of talent flows throughout the company, forming the talent pipeline as a priority, not the talent pool. | Iles, Preece, Chuai (2010), p.127 in Armstrong (2012), p. 256 |
| Using the basic term „talent" to define talent management aims: (a) talent: „what people have when they possess the skills, abilities and aptitudes that enable them to perform effectively in their roles. They make a difference to organizational performance through their immediate efforts and they have the potential to make an important contribution in the future". (b) a talent management aim, generated at this base: It „aims to identify, obtain, keep and develop those talented people". | For (a&b) Armstrong (2012), p. 257; |
| Applying examplary employee talents with a stress on their uniqueness as drivers for support and further elaborations of other researchers' perspectives on HRM/ HCM issues and challenges:<br>(a) generating a list of individual employee talents - superior performance, productivity, flexibility, innovation, and the ability to deliver high levels of personal customer service, organization's competitive position, managing the pivotal interdependencies across functional activities and the important external relationships.<br>(b) formulating a company goal in the HRM/ HCM sphere from the perspective of the resource-based view: to „create more intelligent and flexible firms in comparison to existing competitors by hiring and developing more talented staff and by extending their skills base.<br>(c) providing an analysis through the lens of companies, selling ideas and relationships where knowledge as a direct competitive advantage drives the main challenge to these entities - to „ensure that they have the capability to find, assimilate, compensate and retain the talented individuals they need". | For (a) Armstrong (2011), p. 53<br>For (b) (Boxall, 1996) p. 66 in Armstrong (2011), p. 54<br>For (c) Ulrich (1998) p. 126 in Armstrong (2011), p. 54 |

Table 3. The facets of talent management, outlined by Michael Armstrong (cont'd)

| Specific facet | Source |
|---|---|
| Commenting on a widespread misinterpretation of a basic talent management definition, e.g. talent management is not oriented only to „highflyers", because having better talent at all the levels in the organization is a precondition for gaining a competitive advantage. | Michaels, Handfield-Jones, Axelrod (2001) in Armstrong (2012), p. 257 |
| By defining three main perspectives in choosing the scope of managerial impact on people within talent management efforts: (1) Exclusive people – key people with high performance and/or potential irrespective of position; (2) Exclusive position – the right people in the strategically critical jobs; (3) Inclusive people – everyone in the organization is seen as actually or potentially talented, given opportunity and direction. | Iles and Preece (2010: 248) in Armstrong (2012), p. 257 |
| Describing talent management as a written deliverable, i.e. a specific deliberate strategy, related to this aspect of HRM and in congruence with broad statements of intent in the HRM sphere and overall HR strategies, concerned with high performance working, high commitment management or high involvement management. | Armstrong (2011), p. 125 |

*In a very succinct manner Roberto Luna-Arocas (2012) constructs his collection of shades of meaning for the term 'talent management', accumulated in the years*, and in this way justifies the formulation of his own definition (see table 4). He considers that talent management stems from the sub-field of strategic human resource management (SHRM), bearing three specific characteristics: (a) the existence of a stronger relationship with business strategy; (b) deliberate implementation of different HRM practices for creative pursueing of the same goals as SHRM, and (c) simultaneous application of the configurational approach to SHRM and the system theory as an effective advertisement to attract and sustain the attention of the practitioners.

Table 4. The shades of meaning for talent management according to Luna-Arocas

| Shades of meaning | Source |
|---|---|
| Informal individually-focused talent management | (Tansley, Turner, Foster, 2007) |
| A relabeling of human resource planning | (Lewis, Heckman, 2006) |
| Succession management | (Hirsch, 2000) |
| Strategic character | (Cappelli, 2008; Zuboff, 1998; Boudreau, Ramstad, 2005). |
| Source: Luna-Arocas (2012). | |

*In their quest to provide an overview of corporate experiences with e-learning/e-training and outline appropriate ways of transferring them to the academic institutions (i.e. informtion technology perspctive) Han, Dick, Case, Van Slyke (2012) reflect on the advantages of implementation of Human capital management systems (HCMS) for improving the overall quality of the organization's workforce.* That is why they recommend the use of an integrated strategic HCMS, proposed by Israealite and Seymour (2006) as a means of holistically managing and optimizing the personnel in an organization (see figure 1). In this way talent management comes into being in a specific way as HCMS's key component, fully integrated with the other two ones – learning management and performance management. This component encompasses activities in three sub-spheres

as retention and promotion of top-performing people, their motivating and incenting, and the recruitment and onboarding of high-calibre candidates by the organization. The realizations of the undertaken activities in the last two sub-spheres are supposed to be shared to a different extent with one of the other two components that is emedded in the applied IT solution. The strategic character of talent management is indirectly implied, because ‚strategic' represents an attribute, attached to the label of the aforementioned system.

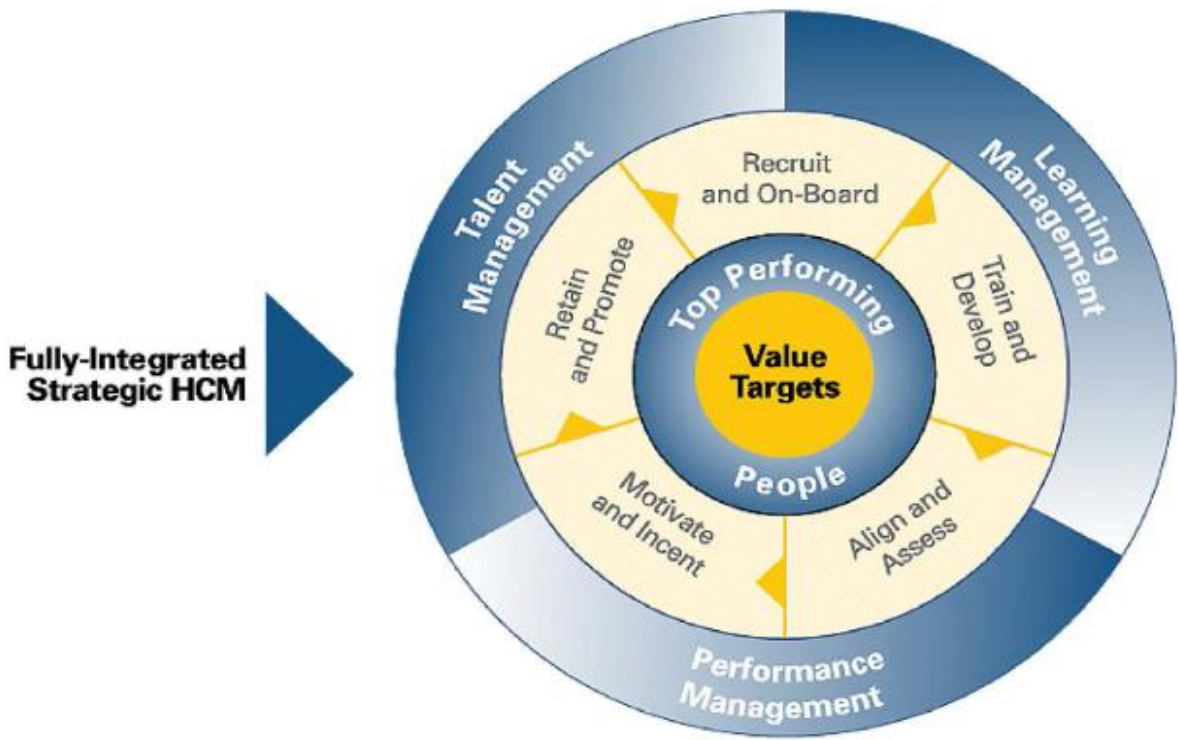

(source: Israealite, Seymour, 2006)

Figure 1. Integrated strategic HCM system

*Ashton and Morton (2005) prefer the use of the case-study approach in order to identify key nuances in the essence of ‚talent' and ‚talent management' in the professional field and research stream of management*. The consultants share the opinion, expressed by Han, Dick, Case, Van Slyke (2012) about the strategic character and holisitcity of talent management, but deliberately extend the scope of its mileau, denoting it as an approach to performing human resource planning, business planning or managers' running a new route to organizational effectiveness. Several nuances in the meaning of talent management, outlined by Ashton and Morton (2005), are summarized in table 5.

Table 5. Assumptions of Ashton and Morton (2005) for the essence of talent management

| Asumptions | Description |
|---|---|
| The need of providing an empirical evidence for talent management realizations in real companies | Presentation of definitions for the base term „talent", formulated in two case companies:<br>(a) Executive management team leaders, directors/VPs and A-player managers in all functions – plus B-players as potentials.<br>(b) Future business leaders with more strategic capabilities than just operational excellence skills – plus specialist talent able to execute business integration projects on time and to budget. |
| Existence of dynamics of „talent" definition in time and respective stages of company life-cycle. | Clearly, there isn't a single consistent or concise definition. Current or historic cultural attributes may play a part in defining talent, as will more egalitarian business models. Many organizations acknowledge that talent, if aligned with business strategy – or the operational parameters of strategy execution – will change in definition as strategic priorities change. For example, in start-up businesses, the talent emphasis will be different to the innovative or creative talent needed to bring new products to market. Any definition needs to be fluid – as business drivers change, so will the definitions of talent. |
| There is a linguistic perspective in thinking about talent management | A list of key words and related explanations is created:<br>(a) *Ethos* – embedding values and behavior, known as a "talent mindset," to support the view that everyone has potential worth developing.<br>(b) *Focus* – knowing which jobs make a difference and making sure that the right people hold those jobs at the right time.<br>(c) *Positioning* – starting at the top of the organization and cascading throughout the management levels to make this a management, not HR, initiative.<br>(d) *Structure* – creating tools, processes and techniques with defined accountability to ensure that the work gets done.<br>(e) *System* – facilitating a long-term and holistic approach to generate change. |
| The necessity of establishing a strategic balance between manager's/ employee performance and potential. | (a) Performance – the primary focus of its measurement and management concerns both the past and the present.<br>(b) Potential - represents the future. It exists, can be identified and developed at all levels in the personnel, so each member may reach his/her potential, no matter what that might be[3]. |
| Source: Ashton, Morton (2005). | |

A steadfast support to the strategic character of talent management is also provided by Hatum (2010, p. 13) who describes it as „a strategic activity aligned with the firm's business strategy that aims to attract, develop, and retain talented employees at each level of the organization. The talent-planning process, therefore, is linked directly to a firm's business and strategic-planning processes".

---

[3] Hatum (2010, p. 16) even lists and defines potential categories of personnel, outlining the scope of talent management in the organization, as follows: (a) top team, (b) middle managers, (c) managers (supervizors), (d) employees, (e) solid performers, (f) critical talent, and (g) high potentials.

In their turn, *Scullion and Collings (2011) determine globalization as the main factor in the formation of talent management meaning. That is why researchers formulated a derivative term, labeled as „global talent management".* It is defined as a rich bundle of „organizational activities for the purpose of attracting, selecting, developing, and retaining the best employees in the most strategic roles (those roles necessary to achieve organizational strategic priorities) on a global scale". The scientists explain the observed dynamics in time and diversity in meaning(s) for this term in global corporations by the impacts of two additional factors, i.e. the specifics in both organizations' global strategic priorities as well as the national contexts for how talent should be managed in the countries where they function.

*Lewis and Heckman (2006) conduct a deeper talent management literature review and devote a whole article in their search for finding an answer to a reasonably posed scientific question: „But what is talent management and what basis does it have in scientific principles of human resources and management?", i.e. following a problematic approach to clarifying shades of meaning for talent management*. Their collection of meanings, arranged by applied criteria, is shown in table 6.

*The last respectively stable and interesting approach to clearing the essence of talent management is to include it in a designed framework of the evolutionary development stages in time for the HR function*. It is supported by both consultancy sector and acadimic field. That is why two close nuances in the aforementioned approach may be identified, as follows:

- The leading consulting businesses rely on a dual facet view to define talent management – the HRM function evolution as a main perspective and the process view (especially, business process management) as a secondary one, ensuring the high-quality servicing for their clients. It permits labeling talent management as „one of the most important buzzwords in Corporate HR and Training today" and is attributed to the current (the third) stage in HR function development (Bersin, 2006). The principal of this leading consulting company (Bersin & Associates) escapes from generating a direct and clear-cult definition for talent management, but unravels its meaning by listing a new set of strategic issues for the organizations in the spheres of HRM, and learning and development (see figure 2). The managers' quest to answering to these questions justifies the design and implementation of „new processes and systems, tigher integration between the different HR silos, and direct..." (real-time) „...integration into line of business management processes" in the business organizations (Bersin, 2006).

Table 6. A collection of meanings for the term ‚talent management' by Lewis and Heckman (2006)

| Criteria | Description | Disadvatages |
|---|---|---|
| Assuming a linguistic perspective | Payng attention to the widespread situation in which researchers and practitioners define „talent management" as a synonymous construct to other (in)stable terms in the sphere of HRM, i.e. „talent strategy", "succession planning\ management", and "human resource planning" | Ambiguity and confusion of outcomes with processes and decision alternatives while reviewing different definitions. |
| Arbitary selection of different definitions for "talent management" from the practitioner-oriented literature | (a) Defined as "a mindset" (Creelman, 2004, p. 3);<br>(b) Defined as a key component to effective succession planning (Cheloha, Swain, 2005);<br>(c) Defined as an attempt to ensure that "everyone at all levels works to the top of their potential" (Redford, 2005, p. 20).<br>(d) Detected perseverance in failures to define the term (Frank, Taylor, 2004; Vicere, 2005; ***, 2005; Ashton, Morton, 2005, p. 30). | The reported picture of talent management is not detailed. |
| The uncovering of three distinct strains of thought regarding talent management[4] | 1. Talent management as a collection of typical human resource department practices, functions, activities or specialist areas such as recruiting, selection, development, and career and succession management. It requires doing what HR has always done but doing it faster (via the internet or outsourcing) or across the enterprise (rather than within a department or function). There are two views to talent management here: (a) with a broad perspective, and (b) prescribing a narrower meaning in comparison to HRM.<br>The tradition is replaced by modernity, i.e.HRM – by talent management. | It cannot provoke deep changes in the principles underlying good recruiting and selection. Its purpose is to re-brand HR practices in order to keep them seemingly new and fresh, without advancing in our knowledge of the strategic and effective management of talent. |
| | 2. Talent management is a set of processes designed to ensure an adequate flow of employees into jobs throughout the organization, i.e. the deliberate formation of talent pools. It is quite close to succession planning/management or human resource planning, recruiting and selection. | Offering just incremental advances in succession management techniques or a closer integration with the organizational staffing models developed in the management sciences. |

---

[4] The same approach is undertaken by Hatum (2010, pp11-12) who constructs his own classification of research streams and respective contents, based on performed literature review for authors'focus of analysis.

Table 6. A collection of meanings for the term ‚talent management' by Lewis and Heckman (2006) (cont'd)

| Criteria | Description | Disadvatages |
|---|---|---|
| The uncovering of three distinct strains of thought regarding talent management[5] | 3. It focuses on talent generically, i.e. without regard for organizational boundaries or specific positions. There are two general views on talent within this perspective:<br>(a) The first one regards talent (which typically means high performing and high potential talent) as an unqualified good and a resource to be managed primarily according to performance levels (A-players: top performers, saught for promotions; B-players: competent performers; C-players: bottom performers, subjected to termination).<br>(b) The second one regards talent as an undifferentiated good and emerges from the both the humanistic and demographic perspectives. The importance of talent is due to two factors:<br>- it is the role of a strong HR function to manage everyone to high performance<br>- demographic and business trends make talent in general more valuable. | Dealing only with programs and processes makes it almost impossible for HR staff to influence the talent inherent in each person, i.e. working with one individual at a time. Managing the "talent inherent in each person" is not a strategic intent. It may be used as a convenient excuse to get rid of low performers. The fact that the organization may not need top performers in all functions as a part of its competitive strategy may be neglected. |

Source: Lewis and Heckman (2006).

- The academic support to this approach applies a more sophisticated framework of HRM function evolution, but limits its interests to alloting undertaken initiatives to exerting desired impacts on (individual) talent and talent planning/ sourcing in the organizations (Ulrich, Younger, Brockbank, Ulrich, 2012)[6]. Mentioning of talent related terms is observed in the text body, describing concisely the essence of some of the proposed four stages of HR work development (i.e. „waves") – HR administration (the first wave), HR strategy (the third wave) and HR outside in (the last forth wave). Here, the authors assume the occurrence of simultaneous realizations of the four waves at the current moment, i.e. enrichment of HR work, although it is implied that numbering sequence denotes earlier emergence in time for waves with lower numbers. Furthermore, up to the moment these stages are predetermined for eternal life, because the shown life-cycle of each wave contains only the stages as start-up, learning, growth and stability. No decline or vanishing of wave-related norms and beliefs about the nature of performed HR work is mentioned. On the contrary alternatives for renewal in conduct by the rules of these waves are outlined. This distribution of talent-related terms across separate stages of HR work evolution does not permit relating talent management theory and practice development with certain key marker events, especially the denoted as a birth year for the term (i.e. 1998) that hints at authors' subconscious assuming of its essence existence without the respective linguistic label in the minds of managers from the business organizations. The emergence in time and the essence of the aforementioned waves is described in table 7 and depicted on figure 3 (A greater emphasis in the table is put on the fourth wave, because of its newness and critical impact on successful market performance of the contemporary companies).

---

[5] The same approach is undertaken by Hatum (2010, pp11-12) who constructs his own classification of research streams and respective contents, based on performed literature review for authors' focus of analysis.
[6] Hatum (2010, p. 21) offers a simplified version of this evolution with just three stages, each one surrounded by its associated HRM activities.

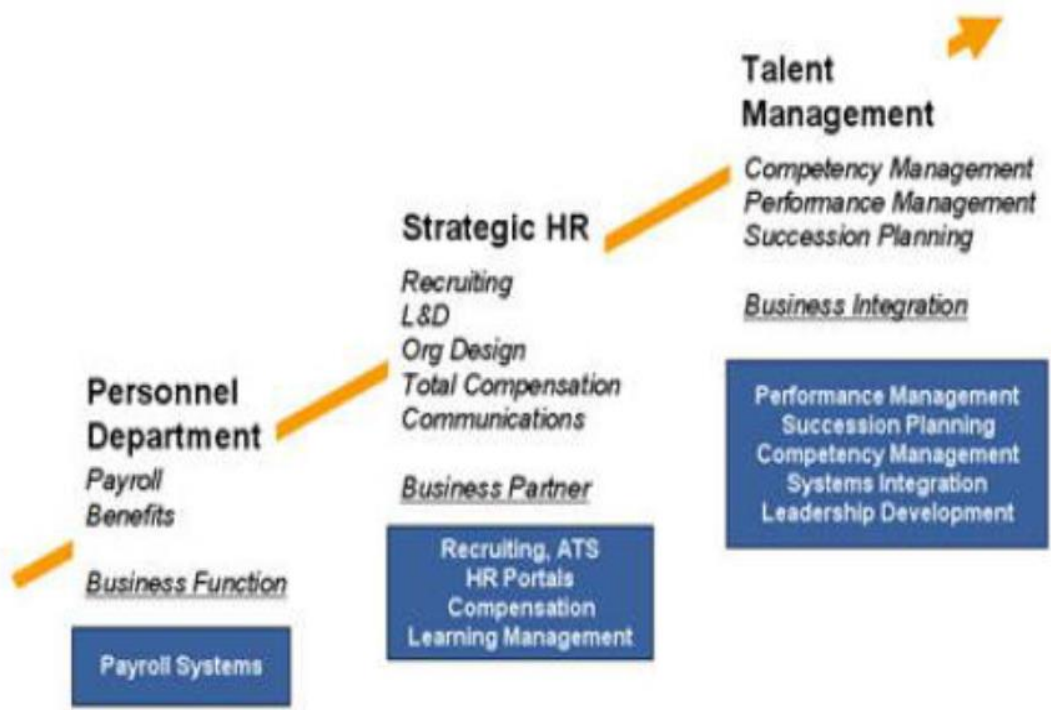

Figure 2. Josh Berin's elucidation on talent management

Table 7. Talent concept and different evolutionary stages in HR work

| Waves of HR work evolution | Description with key takeaways |
|---|---|
| HR administration | The image of the ideal HRs - people who do an excellent job of administration.<br>The primary accountability of HR departments is administrative and transactional. It is maintained nowadays by outsourcing routine work and implementing technology solutions.<br>Training employees, auditing employee satisfaction and engagement, supporting talent planning are characterized as „other important contributions" of the HRs. |
| HR practices | Its oriented to the design of innovative HR practices in sourcing, compensation or rewards, learning, communication, and so forth.<br>Putting an emphasis in integration and consistency among applied practices.<br>Pursuing HR's credibility through delivering of „best practices". |
| HR strategy | A focus on the connection of individual and integrated HR practices with business success through strategic HR (the recent 20 years).<br>Expansion of HR practices from the primary focus on assessing and improving talent to include contribution to culture and leadership to accomplish the pursued business strategy.<br>An enphasis on the link between business strategy and HR actions, and HR credibility that stems from HR's presence „at the table to engage in strategic conversations". |
| HR outside in | Deliberate uses of HR practices to derive and respond to external business conditions.<br>Stretching of the prefessional aspirations beyond strategy to align HR's work with business contexts and stakeholders (for instance conducting 720° performance reviews, clients determine some portion of the bonus pool, etc.).<br>HR's becoming a strategic positioner who knows the business, and can shape and position the business for success.<br>HR's becoming a credible activist who earns personal credibility and also takes an active position on business performance.<br>HR's becoming a capability builder who can find the right mix of personal and organization development actions. Efforts on emphasizing talent are needed.<br>Detected interchangeable use of terms as talent, human capital, workforce, or people.<br>HR's becoming an HR innovator and integrator who weaves separate events into cohesive solutions.<br>HR's becoming an HR change champion who connects the past to the future and who anticipates and manages individual, initiative, and institutional change.<br>HR's using technology to flawlessly process administrative work while generating information for more strategic work. |

Source: (Ulrich, Younger, Brockbank, Ulrich, 2012).

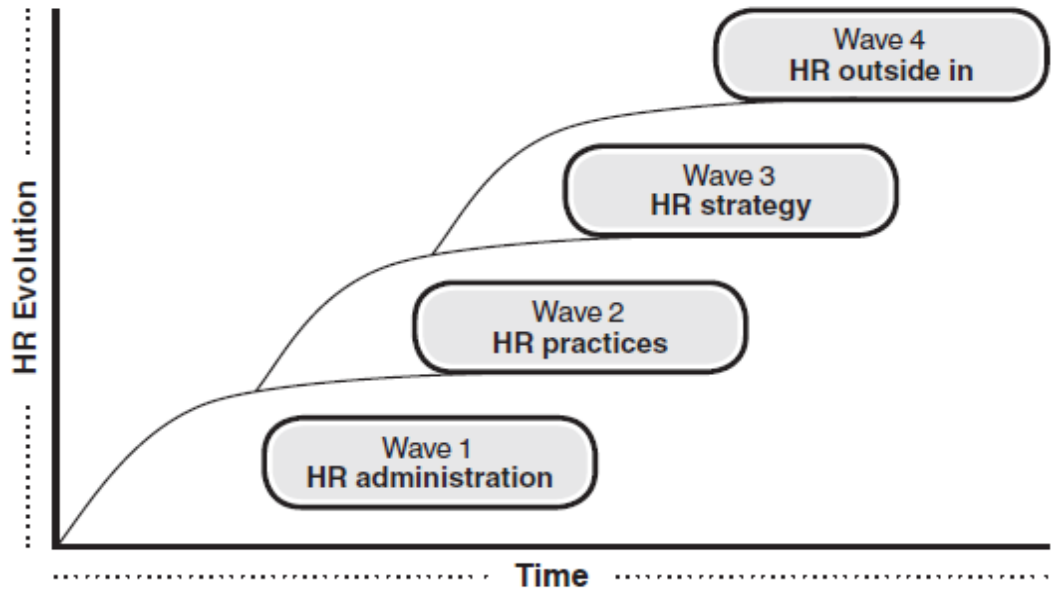

Source: (Ulrich, Younger, Brockbank, Ulrich, 2012).
Figure 3. Evoluion of HR work in waves

### 3. Emerging nuances in the meaning of talent management

*The multi-perspective approach to talent management provides the opportunity of incorporating a variety of managerial approaches to constructing and arranging nuances in the meaning of the explored term*. Thus, Janson (2015) reserves her right to use freely, simulataneously and in congruence with each other a pile of related management terms in order to incorporate what is considered appropiate by a leading HRM consultant, deliberately enriching the meaning of „talent management". In this way she achieves three targets:

- Succeeds in fomulating a more general definition of talent management, revealing it as a mix of „the processes, practices, and activities that are used in hiring people, determining their compensation, managing their job performance, training and developing them, and planning for replacing them should they leave or be promoted".
- Creates a detailed glossary of the most frequently used terms, related to ‚talent managemnt' realizations in business organizations, thus filling it with definite content (see table 8).
- Inherently proposes the idea that taking care of employee is obligatory for contemporary succeeding organizations, but implying that employers do not possess control over employees' potential decisions to undertake career changes, leading them out of the organization.

That is way Janson (2015) finds ways of effective cohesive interpreting the business goals, strategic plan of the company, the assigned team goals and past performance plans into performnace requirements to each individual, determined by role profiles and specific personal goals. These efforts do not divert her attention from individual's professional future, incarnated in design and implementation of individual development plans and respective career plans. The core of management efforts in this sub-sphere of HRM seems to represent the daily activities, associated with employee coaching and feedback as a main mechanism, directing performance reviews and ratings, performed renewal or incremental changes in individual development plans, and remuneration formation.

Table 8. Most widespread talent management terms among practicioners.

| | | |
|---|---|---|
| • 360s | • Feedback | • Performance Management |
| • Banding | • Goals | • Performance Reviews |
| • Bonuses | • Hiring | • Promotions |
| • Career Development | • Individual Development Plans | • Ratings |
| • Career Plans | • Job Evaluations | • Role Profiles |
| • Coaching | • Leadership Models | • Self-Evaluations |
| • Compensation | • Merit Increases | • Stock Options |
| • Competencies | • Needs Assessments | • Succession Planning |
| • Exempt vs. Nonexempt | • Peer Evaluations | • Values |

Source: (Janson, 2015).

Collings and Mellahi (2009, p. 304) also seem to be keen proponents of the multi-facet approach to talent management, in this way providing a very detailed definition for the term as "activities and processes that involve the systematic identification of key positions which differentially contribute to the organization's sustainable competitive advantage, the development of a talent pool of high potential and high performing incumbents to fill these roles, and the development of a differentiated human resource architecture to facilitate filling these positions with competent incumbents and to ensure their continued commitment to the organization".

*Hatum (2010) explores the meaning of talent management by deliberate outlining of roles and responsibilities in this sphere for the respective decision-makers whose integrated efforts may bring in and keep the success in the company*. He expresses his conviction that human resource managers and specialists are not the only owners of talent management activities in the business organization, but the long-term occupation of leading position by the company requires their joint efforts with top managers[7] and line (functional) managers (see figure 4). Of course, some attention is devoted to the on-going and heated discussion around the issues of participation and power distribution among the active stakeholders in this process, proceeding in the business organizations (see also: Stahl, Björkman, Farndale, Morris, Paauwe, Stiles, Trevor, Wright, 2007, 2012; Guthridge, Komm, Lawson, 2008; Collings and Mellahi 2009; Farndale, Scullion, Sparrow, 2010). An important constituency is missed in this analysis, i.e. the employee role (for details see: Ready, Conger 2007; Garrow, Hirsh

---

[7] See (***, 2015).

2008; Stahl, Björkman, Farndale, Morris, Paauwe, Stiles, Trevor, Wright, 2012) and the responsibility of the employee for his or her own career and development is not made more explicit.

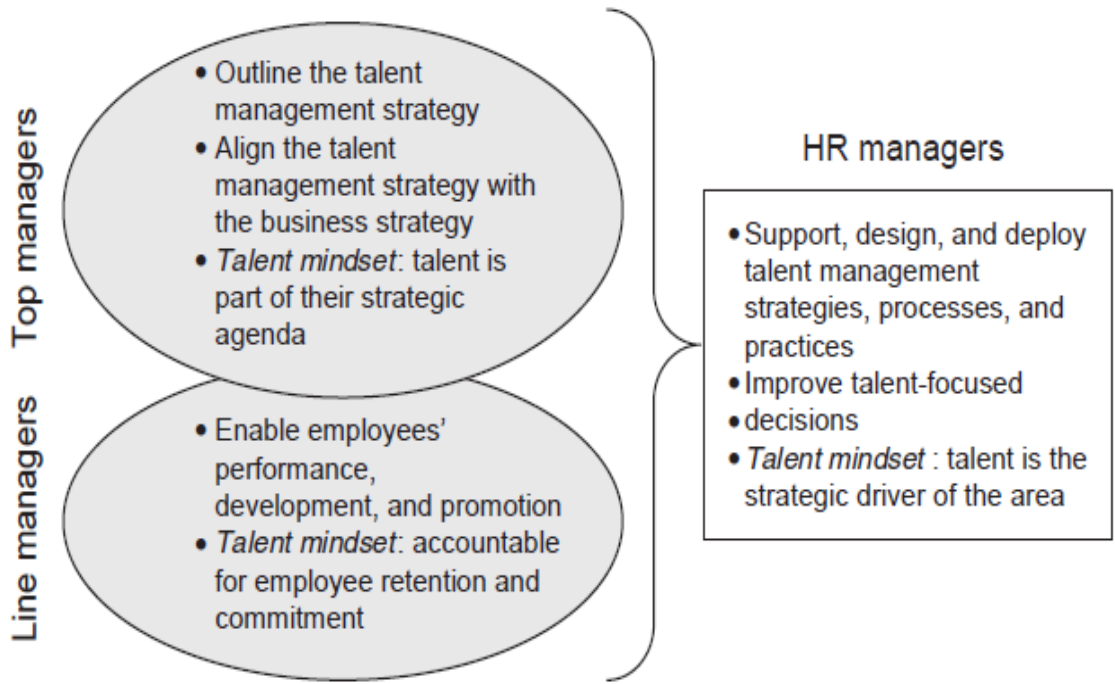

Source: (Hatum, 2010, p. 20).

Figure 4. Roles and responsibilities in talent management

*Exploring the nuances in the meaning of talent management may be accomplished as a result of conducting a survey with a wider main research objective that was the case with Thunnissen, Boselie, Fruytier (2013).* The researchers planned to review and classify „talent management" related literature within the period between 2001-2012 in order to identify and describe dominant themes, leading points of view and delineate omissions, although honestly confessed they could not guarantee having passed through the whole body of scientific literature (i.e. conference papers, dissertations, whole books, book chapters and articles), found in deliberately selected academic electronic databases (i.e. Academic Search Premier, Science Direct, Web of Knowledge and Scopus). They explore the nature and focus of publications and comment that only one third of the items in their literature study include results from empirical research. The team identify the existence of a dominant research assumption among the scientists, working in the sphere of talent management – perceiving it as a transformation process (input, process and output) that involves the usual use of „talent(s) as input, 'process' and develop it (them) with HR practices in order to get the desired output" (Thunnissen, Boselie, Fruytier, 2013). On this basis they outline the emergence of three main issues in publications, oriented to talent management (see table 9).

Table 9. Thematic differentiation of talent management in the past and for the future by Thunnissen, Boselie, Fruytier (2013).

| Themes in talent management | Description | Recommendations for future research and practice of talent management |
|---|---|---|
| The definition of talent | There is no unanimous definition of talent.<br>Several questions have not received unanimous answers by scientists:<br>(a) Whether or not to differentiate the workforce (inclusive or exclusive approach)?<br>(b) What is the appropriate basis for differentiation, or do they assume that the person (subject approach) or competencies (object approach) have to meet the requirements of the organization?<br>There is observed a general agreement on:<br>(a) the impact of the context on the exact and precise description of talent;<br>(b) relativeness and subjectivity of talent, i.e. the mix of differentiating competencies and abilities varies according to the organizational environment (e.g. sector, labor market, customer orientation), the type of work, the internal and external circumstances of an organization and across time. | 1. Elaborating the framework of talent management through surveying in different contexts (for example in different branches of industry, in public, non-profit or private organizations, in small, medium-sized and big enterprises, or national companies versus multinationals, talented people versus non-talented ones and drop-outs etc.).<br>2. Creation of a common language for talent management to strengthen its theoretical foundation by building on and integrating HRM and organizational theories (for example the AMO-model, resource based view, human capital, HR Architecture, career management, contingency theories, socio-technical systems theory, decision-making theories, etc.).<br>3. Emphasizing the empirical research in the sphere of talent management.<br>4. Moving beyond managerialist and unitarist orientation in the talent management literature by acknowledging the existence of clashes of interests, views and goals among numerous constituencies (the organization, its senior or middle managers, supervisors, HRs, employees, colleagues, peers, society, etc.) in its processes and activities (stakeholder theory). |
| Intended effects and outcomes of talent management | Various levels of output and effects are detected: the individual level, the level of the HR-subsystem and the organization as a whole.<br>The scientists are not unanimous on the intended objective by organizations, i.e. earning profit, gaining a competitive advantage or orienting to sustainability.<br>But firm performance is always determined as the main objective of talent management. Firm performance may be influenced by increased employee well-being. | |

Table 9. Thematic differentiation of talent management in the past and for the future by Thunnissen, Boselie, Fruytier (2013) (cont'd).

| Themes in talent management | Description | Recommendations for future research and practice of talent management |
|---|---|---|
| Talent management practices and activities | The attraction, development and retention of talent are the dominant practices and activities in the talent management approaches. Common HR practices and activities are now applied to the field of talent management or to the management of excellence and talent. Context matters. There is no need to prescribe specific practices, but a 'best fit model' is promoted. | 5. Providing an extended consideration of talent management practices and activities (attraction, development and retention of talents, discharge, turnover or moving beyond HRM through work design practices, communication, culture, and leadership). 6. Advocating a greater awareness of contextual fit, beyond the usual focus on strategic or cultural fit[8] 7. Acknowledging the existence of multiple goals for talent management even beyond the entity as societal well-being. |
| Source: Thunnissen, Boselie, Fruytier (2013). | | |

## 4. Discussion and conclusions

At this stage of „talent management" term elaboration the dominating approach among scholars and practitioners is to heap a pile of close, overlapping or supplemental meanings for the continuously enriched construct in text form and/or by insufficient use of some graphical images to depict its current state in order to boost the creativity of opinion leaders to further develop it. An effective tool to perform such a task is the application of a mindmap for the purpose of outlining the formed/ forming perspectives in the essence of talent management. Thus it becomes possible to snapshoot the achievements up to the moment and to delineate direction of potential research and experiments in organizations. Furthermore, innovative ideas and creative solutions of the people, working in the sphere of talent management will be stimulated (see figure 5).

Based on the performed literature review and constructed mindmap a new definition of talent management is proposed in this article, presenting it as *a specific bundle of organization-wide integrated efforts to innovative ideas and creative realizations of contemporary people management that reach far beyond entity's boundaries, balancing diverse interests of firm's constituencies and deliberately searching for their contribution to the process of sustainable value creation not only in the company, but also by integrating its endeavours with other social actors, representing even higher-rank systems, oriented to societal well-being.*

---

[8] Boselie (2010) distinguishes four types of fit: (1) a fit with the organization's strategy (strategic/vertical fit); (2) a fit between individual HR practices (internal/horizontal fit); (3) a fit between the HR strategy and other organizational systems, such as the production system, communication and information system, financial system and legal system (organizational fit) and (4) the link between the human resource strategy and the institutional environment of an organization (environmental fit).

Another way of graphically depicitng talent management essence is by means of a fishbone diagram, outlining the main factors in its essence formation together with the underlying reasons, expressed or inherently implyed by respective researchers.

Furthermore, by contrasting the tradition against the new perspectives in revealing or adding shades of meaning for talent management, it is possible to set new and higher standards of what is permissible, unacceptable, desired, forbidden, verisimilar, or veracious when thinking, defining or working in the field of talent management. That is why some talent management assumptions may be classified as outdated and are respectively labeled as „bad practices". *In this way a contemporary role profile of the underperforming decision-maker in the field of talent management is created, characterizing him as a person who*:

- Demostrates a lack of discipline in applied professional language in the sphere of talent management.
- Does not continuouly search for interweavings of talent management practices with other organization theories and still confines his undertaken interventions within the traditional HRM sphere.
- Does not establish a strategic balance between managers'/ employee performance and potential in talent management conduct.
- Does not understand, predict, accept and use stakeholder impacts on talent management activities of the business organization.
- Demonstrates underdeveloped skills and capabilities, and insufficient knowledge in talent management essence and practices.
- Does not acknowledge the multiple goals perspective of talent management.

Finally, the fishbone diagram serves as a means of mitigating numerous critiques to talent management nuances of meaning (see figure 6).

In conclusion the two aforementioned frameworks may provide the curious readers, scientists and practitioners with a simple and clear explanation of construct's structure, any existing or forming relations among its elements, and important aspects of realized interactions with the higher-rank systems. The beginners in the field will be able to accelerate their learning process in relation to talent management that is a huge issue, since many of the managers, researchers and students are already experts in boundary fields. The multiple production of such graphical tools may reveal the dynamics of construct's elaboration by realizations of multiple snapshooting in time or at the occurrence of key events. In summary, this analysis confirms the avilability of great potential of „talent management" for future elaboration in practice and science as organizations continuously confront people-related challenges during their existence.

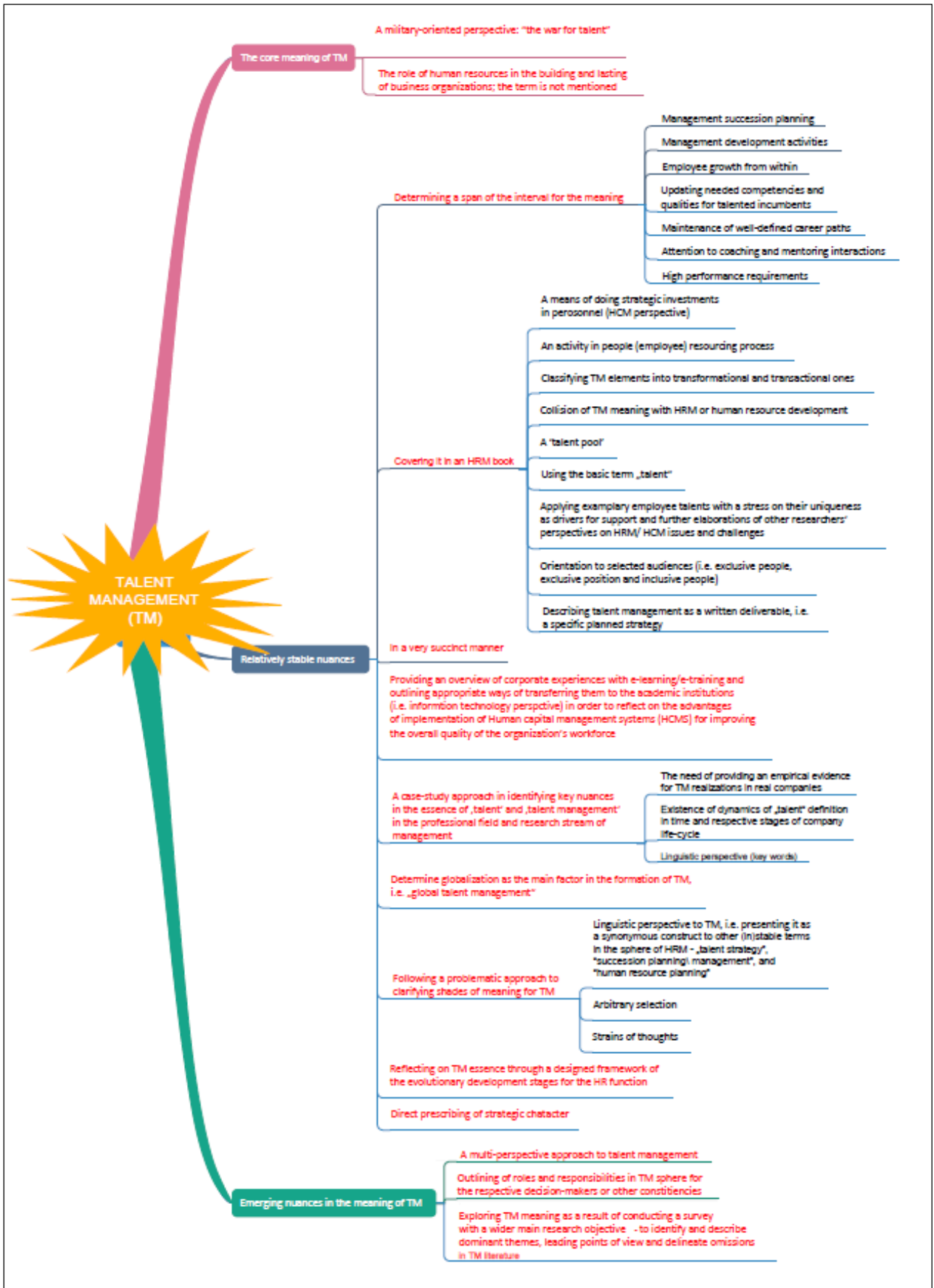

Figure 5. The perspectives in clarifying talent management essence

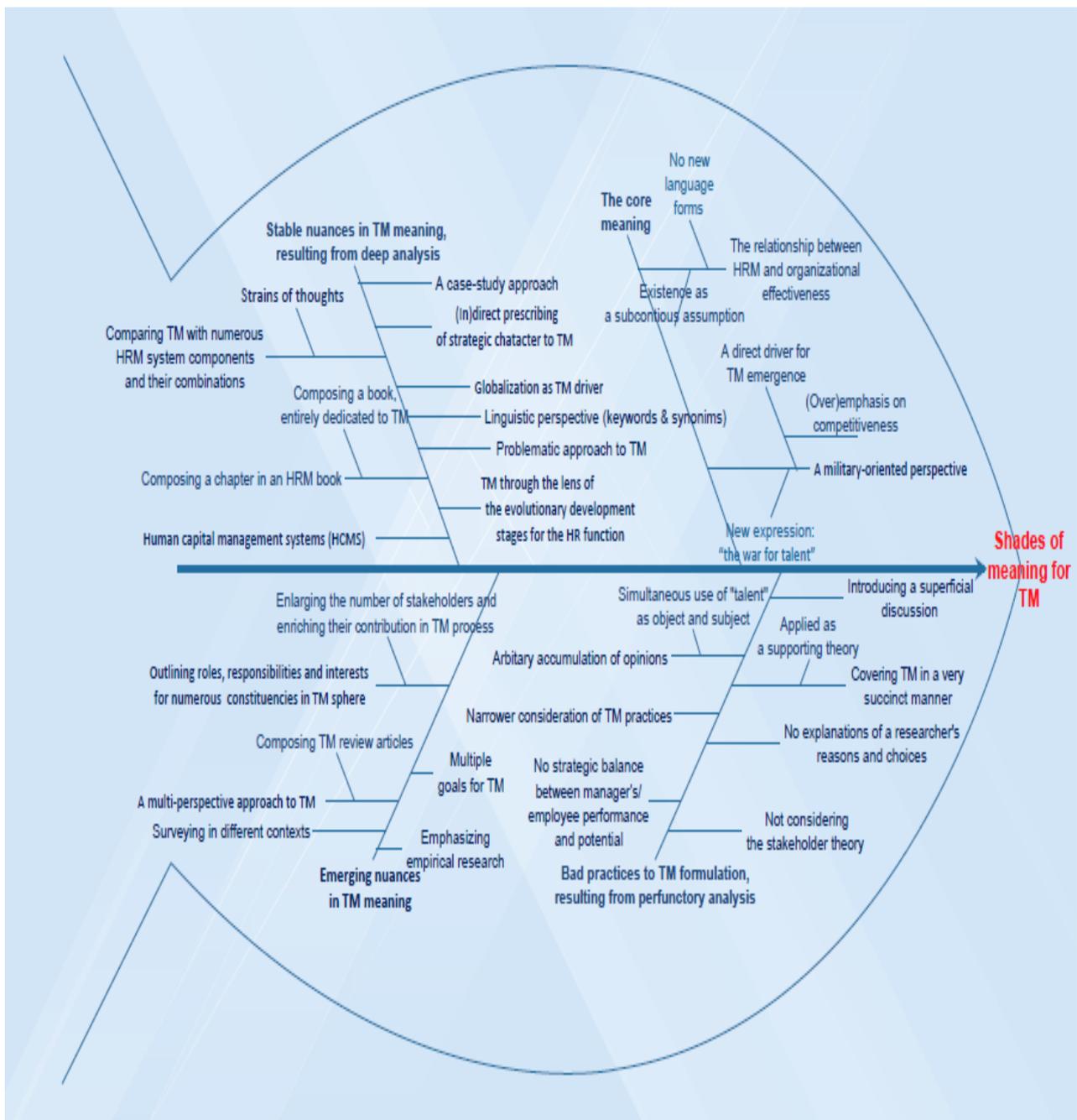

Figure 6. A fishbone diagram of talent management essence

**Information about the author**

associate professor Kiril Dimitrov, Ph.D., "Industrial business" department, University of National and World Economy (UNWE) – Sofia, Bulgaria, e-mail: kscience@unwe.eu

.